\documentclass[useAMS,referee,usenatbib]{biom}
\def\bSig\mathbf{\Sigma}

\usepackage{graphicx}
\usepackage{pdflscape}
\usepackage{rotating}
\usepackage{booktabs}

\title[A mixture model for disease progression subtypes]{A mixture model for subtype identification in the context of disease progression modeling}

\author{Sofia Kaisaridi$^{1}$, 
Juliette Ortholand$^{1,2}$, Caglayan Tuna$^{1,3}$, \\ \textbf{Hugues Chabriat$^{\mathbf{4,5,6}}$ and Sophie Tezenas du Montcel$^{\mathbf{1,*}}$\email{sophie.tezenas@aphp.fr}} \\
$^{1}$ARAMIS, Sorbonne Université, Institut du Cerveau, CNRS, Inria, Inserm,\\ AP-HP, Groupe Hospitalier Sorbonne Université, Paris, France \\
$^{2}$KIK, Amsterdam UMC, Amsterdam, Netherlands\\
$^{3}$Inria, Université Paris Cité, Inserm, HeKA, F-75015, Paris, France\\
$^{4}$GENOVASC, INSERM U1127, Institut du Cerveau, Paris, France\\
$^{5}$Centre de Référence National CERVCO, Paris, France\\
$^{6}$CNVT, FHU-NeuroVasc-2030, Hopital Lariboisiére APHP, Université Paris-Cité, France}

\usepackage{amsfonts, amsmath}
\DeclareMathOperator*{\argmax}{argmax}
\usepackage{derivative}
\usepackage{hyperref}

\begin{document}




\pagerange{\pageref{firstpage}--\pageref{lastpage}} 




\label{firstpage}


\begin{abstract}
The progression of chronic diseases often follows highly variable trajectories, and the underlying factors remain poorly understood. 
Standard mixed-effects models typically represent inter-patient differences as random deviations around a common reference, which may obscure meaningful subgroups.
We propose a probabilistic mixture extension of a mixed effects model, the Disease Course Mapping model, to identify distinct disease progression subtypes within a population. 
The mixture structure is introduced at the latent individual parameters, enabling clustering based on both temporal and spatial variability in disease trajectories.
We evaluated the model through simulation studies to assess classification performance and parameter recovery. 
Classification accuracy exceeded 90\% in simpler scenarios and remained above 80\% in the most complex case, with particularly high recall and precision for fast-progressing clusters. 
Compared to a post hoc classification approach, the proposed model yielded more accurate parameter estimates, smaller biases, lower root mean squared errors, and reduced uncertainty. 
It also correctly recovered the true three-cluster structure in 93\% of the simulations.
Finally, we applied the model to a longitudinal cohort of CADASIL patients, identifying two clinically meaningful clusters, differentiating patients with early versus late onset and fast versus slow progression, with clear spatial patterns across motor and memory scores. 
Overall, this probabilistic mixture framework offers a robust, interpretable approach for clustering patients based on spatiotemporal disease dynamics.
\end{abstract}

%



\maketitle


%

\section{Introduction}\label{intro}

Chronic diseases exhibit complex and highly variable trajectories that unfold over years yet, the factors driving this heterogeneity remain poorly understood. 
Accurately modeling disease progression is essential not only for predicting individual outcomes but also for uncovering pathological mechanisms that can guide targeted interventions.
Mixed-effects models offer a powerful framework for reconstructing disease trajectories while accounting for variability between patients (random effects) and observed variables (fixed effects).
Traditional approaches often assume that inter-patient variability manifests as random deviations around a fixed reference trajectory.
While mathematically convenient, this assumption may overlook structured variability. 
Genetic mutations, environmental factors and demographic characteristics, can systematically shape disease progression, highlighting the limitations of treating variability as random noise~(\cite{Verbeke1996}).
Nevertheless, in many chronic diseases, the mechanisms driving progression and the relevant and baseline covariates remain insufficiently understood, complicating the identification of clinically meaningful patient subgroups. 

Probabilistic mixture models offer a principled way to capture heterogeneity by modeling variability directly at the parameter level and grouping individuals with similar progression patterns~(\cite{Proust2005}). 
Several related approaches exist. 
A latent class mixed effects model estimated via an iterative Marquardt algorithm is implemented in the R library lcmm~(\cite{Proust2017}).
Although it defines a latent structure, it does not explicitly employ a data-driven formulation of disease age~(\cite{Young2024}).
As a result, individuals are compared at similar chronological ages, which may confound heterogeneity in timing with heterogeneity in dynamics.
Finite mixture modeling has also been applied to cross-sectional data through the Subtyping and Staging model implemented in the Python library SuStaIn~(\cite{Young2018}). 
In addition, a mixed-effects model using the MCMC-SAEM algorithm is available in the saemix package~(\cite{Comets2017}); however, to the best of our knowledge, it has not been extended to mixture models.

A non-linear mixed-effects disease progression model, built within a Riemannian geometry framework, named the Disease Course Mapping model~(\cite{Schiratti2015,Schiratti2017}) estimates parameters using an MCMC-SAEM algorithm.
It captures temporal dynamics through time reparametrization and spatial (topological), patterns of progression.
Compared with classic mixed effects models, this geometrical formulation is particularly suited to multivariate settings and provides interpretable insights into the coordinated evolution of multiple clinical scores.
It has already been proven useful in various settings such as constructing disease course maps~(\cite{Koval2021}), predicting individual disease profiles~(\cite{Maheux2023}), modeling of discrete responses~(\cite{Poulet2023, Moulaire2023}), joint modeling of events~(\cite{Ortholand2025_multi},\cite{Ortholand2025_uni}) and handling multiple scores~(\cite{Kaisaridi2025}).
Building on this framework, we introduce a mixture Disease Course Mapping model, designed to identify subgroups with similar disease trajectories.
Unlike earlier work that considered only the temporal aspect~(\cite{Poulet2021}), our approach introduces a unified formulation in which temporal alignment and spatial progression are modeled jointly within a probabilistic mixture framework.
Related mixture models have been proposed for pice-wise curves~(\cite{Debavelaere2020}) where mixture components both population and individual level parameters.
While these approaches allow for rich geometrical descriptions of disease evolution, they typically rely on complex population parameterizations.
Our model operates entirely at the individual level while preserving a shared global disease trajectory, thereby balancing individual heterogeneity with population coherence.
Progression is modeled using a logistic trajectory, providing a parsimonious yet biologically plausible representation of monotonic disease evolution, capturing the typical slow onset, accelerated change, and saturation, without shifts induced by piecewise curves.
We validate our approach on simulated data and demonstrate its advantages over a naive post-hoc classification, in capturing structured heterogeneity and enhancing interpretability and predictive accuracy.
We also illustrate an application using real data from CADASIL (Cerebral Autosomal Dominant Arteriopathy with Sub-cortical Infarcts and Leukoencephalopathy), the most frequent inherited cerebral small vessel disease (cSVD), whose gradual and heterogeneous evolution has been documented previously~(\cite{Brice2020,Brice2022,Kaisaridi2025}).

\section{Model}\label{sec: Model}

\subsection{The Disease Course Mapping model}

The Disease Course mapping model was first introduced as a non-linear mixed effects model with a time reparametrization~(\cite{Schiratti2015,Schiratti2017}).
We consider a dataset comprising $N$ patients, each undergoing $n_i$ clinical visits, during which up to $d$ features are assessed. 
Let $t_{ij}$ denote the age of patient $i$ at visit $j$, and let $y_{ijk}$ represent the value of feature $k$ for patient $i$ at visit $j$. 
In the multivariate logistic formulation of the Disease Course Mapping model, each feature follows a logistic trajectory modeled as:

\begin{eqnarray}
y_{ijk} = \left(1 + (\frac{1}{p_k}-1)exp(-\frac{v_k(e^{\xi_i}(t_{ij}-\tau_i))+w_{ik}}{p_k(1-p_k)})\right)^{-1} +\epsilon_{ijk} \label{eq: DCM_statmodel}
\end{eqnarray}

Following the hierarchical structure commonly employed in mixed effects models, two types of parameters are introduced to capture both population-level and individual-level (random) effects.
The population-level parameters $p_k$ and $v_k$ represent, respectively, the $k$-th coordinate of the position and velocity at the average disease onset time $\overline\tau$.
The individual-level parameters $\xi_i$ and $\tau_i$ characterize temporal variability across subjects: $\xi_i$ represents the individual progression rate, while $\tau_i$ the individual onset time of the disease for subject $i$.
A positive $\xi_i$ indicates a faster progressing profile, and a higher value of $\tau_i$ corresponds to a later disease onset.
The time reparametrization function $e^{\xi_i}(t_{ij}-\tau_i)$ aligns individuals along a common disease timeline, reduces confounding due to heterogeneous observation times, and enables comparisons between subjects at comparable disease stages rather than at similar chronological ages.
The variables $w_{ik}$ capture spatial variability, in a topological rather than geographic sense, and can be interpreted as score-specific space shifts.
A positive $w_{ik}$ indicates that the $k$-th score is more advanced for the $i$-th patient at the time of average disease onset, and conversely for negative values.
They are not estimated directly but are instead decomposed along $N_s$ sources $s_i$ with the help of a mixing matrix in the spirit of an ICA decomposition.
The number of sources $N_s$ defines the degrees of freedom of the spatial variability, corresponding to groups of scores that share similar spatial behavior.
Finally, a Gaussian noise term is added to account for measurement variability $\epsilon_{ijk} \sim \mathcal{N}(\mathbf{O}_d, \sigma_k^2\mathbf{I}_d)$.
A more detailed description of the model can be found in the supplementary material.

\subsection{Mixture Model}

We now assume that multiple subtypes of disease progression may coexist within the same population. 
Each subtype $c$ is characterized by its own set of parameters $(\overline{\tau}^c, \overline{\xi}^c, \overline{s}^c)$ representing the mean onset time, progression rate, and spatial pattern, respectively, and occurs with an associated probability $\pi^c$.

We introduce a probabilistic framework in which each individual $i$ is assumed to belong to a cluster $c$, ($c=1, 2, \dots, n_c$)  with probability $\pi_i^c$.
The model assumes a prespecified number of clusters $n_c$, defined before estimation.
The overall probability of occurrence of cluster $c$ across the population is then defined as $\pi^c = \frac{1}{N} \sum_{i}\pi_i^c$ with $\pi^c \geq 0$ and $\sum_c\pi^c=1$.
To formalize this, we introduce a latent individual variable that encodes the cluster membership, defined as a one-hot encoded vector $\mathbf{r_i}$ where the $c$-th element indicates whether individual $i$ belongs to cluster $c$. 
Each element of the vector takes a binary value, where $r_i^c=1$ if $c=\argmax_c \pi_i^c$ and $r_i^c=0$ otherwise.
Conceptually, this is equivalent to: $r_i^c=1$, if individual $i$ belongs to cluster $c$ and $0$ otherwise.
So then, the number of individuals belonging to each cluster conditioning on the latent indicators is defined as $N_c := \sum_{i=1}^N r_i^c$.

\subsubsection{Estimation}

\paragraph{Parameters}

For estimation purposes, three categories of parameters are defined: latent parameters $z$, model parameters $\theta^c$, and hyperparameters $\Pi$.
Latent parameters are sampled during fitting according to probability distributions defined directly or after transformations.
Specifically, the transformations used are : $g_k = \frac{1 - p_k}{p_k}$, $\tilde{g}_k = \log g_k$, $\tilde{v}_k = \log v_k$.
We distinguish between population parameters $\mathbf{z}_{pop} =(\tilde{g_k}, \tilde{v_k}, \beta_{ml})$ and individual parameters $\mathbf{z}_{re,i} =(\tau_i, \xi_i, s_i)$. 
The population-level parameters are assumed to follow normal distributions, as in the standard Disease Course Mapping model, whereas the individual-level parameters are drawn from a mixture of Gaussians with mixing probabilities $\pi^c$.

\begin{equation}
\begin{minipage}[t]{0.48\linewidth}
\centering
\textbf{Population-level parameters:}
\[
\begin{cases}
    \tilde{g}_k \sim \mathcal{N}(\overline{\tilde{g}_k}, \sigma^2_{\tilde{g}}),\\[4pt]
    \tilde{v}_k \sim \mathcal{N}(\overline{\tilde{v}_k}, \sigma^2_{\tilde{v}}),\\[4pt]
    \beta_{ml} \sim \mathcal{N}(\overline{\beta_{ml}}, \sigma^2_{\beta})
\end{cases}
\]
\end{minipage}
\hfill
\begin{minipage}[t]{0.48\linewidth}
\centering
\textbf{Individual-level parameters:}
\[
\begin{cases}
    \xi_i \sim \sum_{c=1}^{n_c} \pi^c \mathcal{N} (\overline\xi^c, {\sigma_{\xi}^c}^2) \\
    \tau_i \sim \sum_{c=1}^{n_c} \pi^c \mathcal{N} (\overline\tau^c, {\sigma_{\tau}^c}^2)\\
    s_{il} \sim \sum_{c=1}^{n_c} \pi^c \mathcal{N} (\overline s_l^c, {\sigma_{s}^c}^2)
\end{cases}
\]
\end{minipage}
\label{eq : latent_params}
\end{equation}

The model parameters $\theta^c$ are the means and standard deviations of the distributions of all the latent parameters, along with the variance of the Gaussian noise $\epsilon_{ijk}$ added to the observations.
They correspond to the fixed effects.
They are the parameters estimated by the model $\mathbf{\theta}^c =(\overline{\tilde{g}}_k,\overline{\tilde{v}}_k, \overline{\beta}_{ml},\overline{\xi}^c, \sigma_{\xi}^c, \overline{\tau}^c, \sigma_{\tau}^c, \overline{s_l}^c,  \sigma_{\kappa})$.
The standard deviations $\sigma_{\tilde{g}}$,$\sigma_{\tilde{v}}$ $\sigma_{\beta}$ are all fixed to 0.01 by default. 
Intuitively, they provide the exploration strength during fitting for the population parameters.
We also fix $\sigma_{s}^c=1$ consistent with the standard Disease Course Mapping model formulation to reduce model complexity and prevent overparameterization.
In addition, we impose centering constraints on the individual parameters $\xi_i$ and $s_{il}$ across all subjects to ensure the orthogonality conditions of the general model.
The parameters $\{ \sigma_{\tilde{g}}, \sigma_{\tilde{v}}, \sigma_{\beta}, \sigma_s^c \}$ form the set of hyperparameters $\Pi$.

\paragraph{Likelihood}

The likelihood estimated by the model is : $p(y \mid \theta^c, \Pi) = \int_{z} p(y, z \mid \theta^c, \Pi) dz$.
For convenience in estimation, we consider the log-likelihood, and we divide the term $\log p (y, z \mid \theta, \Pi)$ into three different terms. 
The data attachment, 
and two terms for the prior attachment of latent individual and population parameters. 
As the mixture is assumed to interact only with the individual parameters, the components associated with the data and the population parameters remain identical to those defined in the standard Disease Course Mapping model~(\cite{Schiratti2017}) and they are detailed in the supplementary material.
The distribution of the random effects is modeled as a Gaussian mixture conditioned on the latent class indicators $r_i^c$, so the relevant likelihood component takes the form:

\begin{align}
\log p(z_{re} \mid z_{fe}, \theta^c, \Pi)
&= \sum_{i=1}^N \sum_{c=1}^{n_c} r_i^c\Big[\log\pi^c
- \log(\sigma_{z_{re}}^c\sqrt{2\pi})
- \frac{(z_{re,i}-\mu_{z_{re}}^c)^2}{2(\sigma_{z_{re}}^c)^2}\Big].
\label{eq: likelihood}
\end{align}

Introducing the latent indicators $r_i^c$, ensures that the complete model belongs to the curved exponential family, guaranteeing convergence of the MCMC-SAEM algorithm~(\cite{Kuhn2004}). 
While the Gaussian mixture model does not belong to a regular exponential family, since the marginal likelihood involves a log-sum over components, it inherits an exponential family structure conditionally on $r_i^c$. 
Marginalizing over the latent indicators renders the model curved, meaning that the parameters $(\pi^c, \theta^c)$ lie on a nonlinear surface within the full exponential-family space corresponding to all possible assignments of $r_i^c$. 
Consequently, this formulation enables derivation of the sufficient statistics.
Further details and proofs are provided in the supplementary material.

\paragraph{MCMC-SAEM}

The model formulates as a nonlinear mixed-effects model for which the observed likelihood does not admit a closed-form expression. 
Consequently, an analytical maximization of the log-likelihood is not feasible. 
To estimate the model parameters, we employ a stochastic version of the Expectation–Maximization (EM) algorithm, namely the MCMC-SAEM algorithm (Monte Carlo Markov Chain – Stochastic Approximation Expectation Maximization).
The stochastic approximation mechanism ensures convergence for the models that lie in the curved exponential family~(\cite{Kuhn2004}). 
Yet, for the mixture model, we need to adapt the MCMC-SAEM algorithm to ensure that the estimation is performed independently for each cluster. 
We adapted a version known as the mixture MCMC~(\cite{McLachlan2008}) with the SAEM mechanism~(\cite{Lavielle2014}).
This approach iteratively alternates between three main steps:
\begin{enumerate}
    \item a simulation step, where the latent variables $\mathbf{z}^c$ are sampled according to their distributions given the current estimates of the model parameters using a Metropolis-Hastings algorithm with a Gibbs sampler
    \item a stochastic approximation step, where instead of computing the expected complete-data log-likelihood explicitly, its sufficient statistics are updated recursively using a stochastic averaging scheme~(\cite{Robbins1951})
    \item a maximization step, where the model parameters $\boldsymbol{\theta}^c$ for each mixture component are updated by maximizing the approximate expected complete-data log-likelihood
\end{enumerate}

A detailed description of the mixture MCMC-SAEM algorithm and the parameter update rules is provided in the supplementary material.
After convergence, the posterior distribution of the individual latent variables $z_{i}$ can be used to compute individual cluster membership probabilities: $\pi_i^c = \frac{\pi^c \, p(z_{re,i} \mid z_{fe}, \theta^c, \Pi)}{\sum_{c} \pi^{c} \, p(z_{re,i} \mid z_{fe}, \theta^{c}, \Pi)}$.
The MCMC already incorporated the observed data, so this posterior is consistent with the full model.
Hard cluster assignments can then be obtained by selecting the component with the highest posterior probability, as commonly done in finite mixture models~(\cite{Proust2017}).

\subsubsection{Implementation}

The proposed mixture model is implemented in the latest released version of open-source python library leaspy \href{https://github.com/aramis-lab/leaspy/}{https://github.com/aramis-lab/leaspy/}.
The python code used for the simulations and the statistical analysis can be found in this github repository \href{https://github.com/KaisaridiSofia/Mixture_Model_Leaspy}
{https://github.com/KaisaridiSofia/Mixture\_Model\_Leaspy}.

\section{Simulation study}

\subsection{Method}

We followed the ADEMP recommendation~(\cite{Morris2019}) to perform a simulation study in order to validate the proposed model.

\subsubsection{Aims}

The performance of the proposed method was evaluated with respect to two fundamental objectives: the accurate classification of individuals into their corresponding clusters and the precise estimation of the underlying cluster-specific parameters.
In the context of mixture modeling, the first objective corresponds to the model’s ability to correctly assign each individual to its true cluster.
The second objective concerns the model’s capacity to recover, with minimal bias, the parameters characterizing each cluster—specifically, in this study, the mean values of the individual parameters.

\subsubsection{Data generating mechanism}

The data-generating mechanism used in this study was inspired by a recent longitudinal study in CADASIL patients~(\cite{Kaisaridi2025}). 
By analyzing data from patients recruited at the French National Referral Centre CERVCO, this study revealed two profiles of spatiotemporal progression.
An early-onset, rapidly progressing cluster with less advanced focal deficits, motor disability, and dependency at the reference onset age, and a late-onset, slowly progressing cluster with cognitive symptoms that remain less advanced at the reference onset age. 

Three different scenarios were considered.
For the first scenario, we generated a population imitating the CADASIL example by simulating two clusters with the spatiotemporal profiles described above.
For all individuals, we simulated two longitudinal scores, each having a different manifestation for the subgroups revealed in the previous study (\textit{scenario\_2\_2}: two scores and two clusters).
For the second scenario, we simulated two clusters that were more intertwined. 
For that, we included a third score that shows similar progression along the two clusters (\textit{scenario\_3\_2}: three scores and two clusters).
Finally, we included a third scenario (\textit{scenario\_multi}: six scores and three clusters) to examine the performance of the model in more complex settings involving more clusters and scores.
For each simulation scenario, 1,000 independent datasets were generated, each consisting of 1,000 patients with six visits occurring at random time intervals.
For all parameters, individual parameters were generated using normal distributions, with the means in table~\ref{tab: simulation_values} and the following standard errors: $\sigma_{\tau}=5$, $\sigma_{\xi}=0.5$, $\sigma_s=1$.
Data were simulated under the Disease Course Mapping model structure according to the following procedure:

\begin{itemize}
\item Fixed effects were derived from the fit of a mixture Disease Course Mapping model applied to real data from CADASIL patients recruited at the French National Center CERVCO to emulate realistic scenarios.
\item Random effects were simulated independently for each cluster from the corresponding Gaussian distributions.
\item An increasing sequence of visit times was generated for each individual, centered around their onset time $\tau_i$, with randomly spaced gaps and added Gaussian noise to reflect realistic variability.
\item The Disease Course Mapping logistic multivariate formulation was used to compute simulated longitudinal score values for all individuals.
\end{itemize}

\subsubsection{Estimands}

For each dataset, we first fitted a mixture Disease Course Mapping model using the appropriate number of clusters for the given scenario. 
From this model, we obtained both the population-level and individual-level parameters, and the posterior probabilities for each individual assignment. 
To enable a comparison between the \textit{a priori} estimation of the cluster parameters provided by the mixture model, and the \textit{a posteriori} classification, we also fitted a standard Disease Course Mapping model to each dataset, followed by a naive post-hoc clustering procedure. 
with the use of the module GaussianMixture from the sklearn library~(\cite{scikit2011}).

In mixture models, the likelihood function is invariant to permutations of the cluster labels, meaning that if two clusters are swapped, along with their associated parameters, the model likelihood remains unchanged. 
As a result, during estimation, the labeling of clusters can become arbitrary, leading to what is known as the label switching problem. 
In practice, this manifests as estimated cluster centers that do not consistently correspond to the true underlying components.
To correct for label switching, we aligned the estimated cluster centers with the true (simulated) parameters using a distance-based matching procedure.
Estimated and true parameters were first z-score normalized, and a Euclidean distance matrix was computed between them.
The optimal one-to-one assignment minimizing the total distance was then determined using the Hungarian algorithm~(\cite{Kuhn1955}), providing a mapping between estimated and true cluster labels for each simulation.

\subsubsection{Performance metrics}

We first assessed the ability of the Integrated Completed Likelihood (ICL) criterion~(\cite{Biernacki2000}) to correctly identify the number of clusters.
As the model-based clustering criteria, such as ICL tend to favor parsimonious, well-separated clusters rather than overfitting the number of mixture components~(\cite{Biernacki2000}), we chose to test it only in the third scenario with three clusters, rather than in the two-cluster scenarios.
Then, the performance of the proposed model was evaluated in terms of both classification and parameter estimation for all the scenarios.

\paragraph{Determination of the number of clusters}

We used the third simulation scenario (\textit{scenario\_multi}) to examine the behavior of the ICL criterion with respect to cluster-number selection in our simulation setting. 
Datasets were generated under a three-cluster assumption, and mixture models consistent with this specification were initially fitted. 
For comparison, mixture models assuming two and four clusters were also fitted to the same datasets, and the proportion of datasets for which the correct specification (three clusters) was preferred was subsequently assessed.

\paragraph{Classification}

For the assessment of classification performance, we compared the predicted cluster assignments with the true labels using a confusion matrix, from which standard classification metrics were derived.
Accuracy was defined as the proportion of correctly assigned individuals among all observations, quantifying the overall correctness of the clustering.
Recall (or completeness) quantified the proportion of true cluster members that were correctly identified, thereby indicating how well the method recovered each true cluster. 
Precision (or purity) measured the proportion of true positives among all individuals assigned to a given cluster, reflecting how often a predicted cluster corresponded to the true group. 
For all classification metrics, 95\% confidence intervals were also calculated.
Performance estimates were obtained across 1,000 independent simulations, each comprising 1,000 individuals, and the reported values represent the averages across these repetitions.
To disentangle the effects of model selection and patient classification, an additional analysis was conducted on the subset of datasets for which the ICL criterion selected the correct number of clusters for the third scenario.

\paragraph{Parameter estimation}

To evaluate the fit of the estimated model parameters, we compared the estimated parameter values ($\hat{\theta}$), obtained as the mean over 1,000 simulations, with their known true values ($\theta_{\text{true}})$, for the parameters that define the cluster centers temporally $\overline{\tau^c}$, $\overline{\xi^c}$ and the score-specific space shifts $\overline{w_k^c}$ for each score $k=1,\dots, d$.
Model estimation performance was summarized using three standard criteria: the bias, defined as $E(\hat{\theta}) - \theta_{\text{true}}$, representing systematic deviation from the true value, the root mean squared error (RMSE), defined as $\sqrt{E[(\hat{\theta} - \theta_{\text{true}})^2]}$, 
reflecting both bias and variance in estimation, and the standard error (SE), defined as
$\sqrt{Var(\hat{\theta})}$ quantifying the variability of the estimator across replicates.
Together, these metrics provide a comprehensive evaluation of the parameter estimation accuracy of the proposed mixture model.
To disentangle the effects of model selection and parameter estimation, analysis was repeated on the subset of datasets correctly classified by the ICL criterion in terms of cluster number, for the third scenario.

\section{Results}

\subsection{Determination of the number of clusters}

Across 1,000 datasets generated under the three-cluster assumption, the model selected as optimal according to the ICL criterion was the three-cluster model in 93.2\% of cases, the two-cluster model in 5.5\%, and the four-cluster model in 1.3\%. 
The results presented below are based on the full set of 1,000 datasets, including those for which the ICL criterion favored a different number of clusters.

\subsection{Classification}

\begin{figure}[!t]
\centerline{\includegraphics[width=\textwidth,height=46pc]{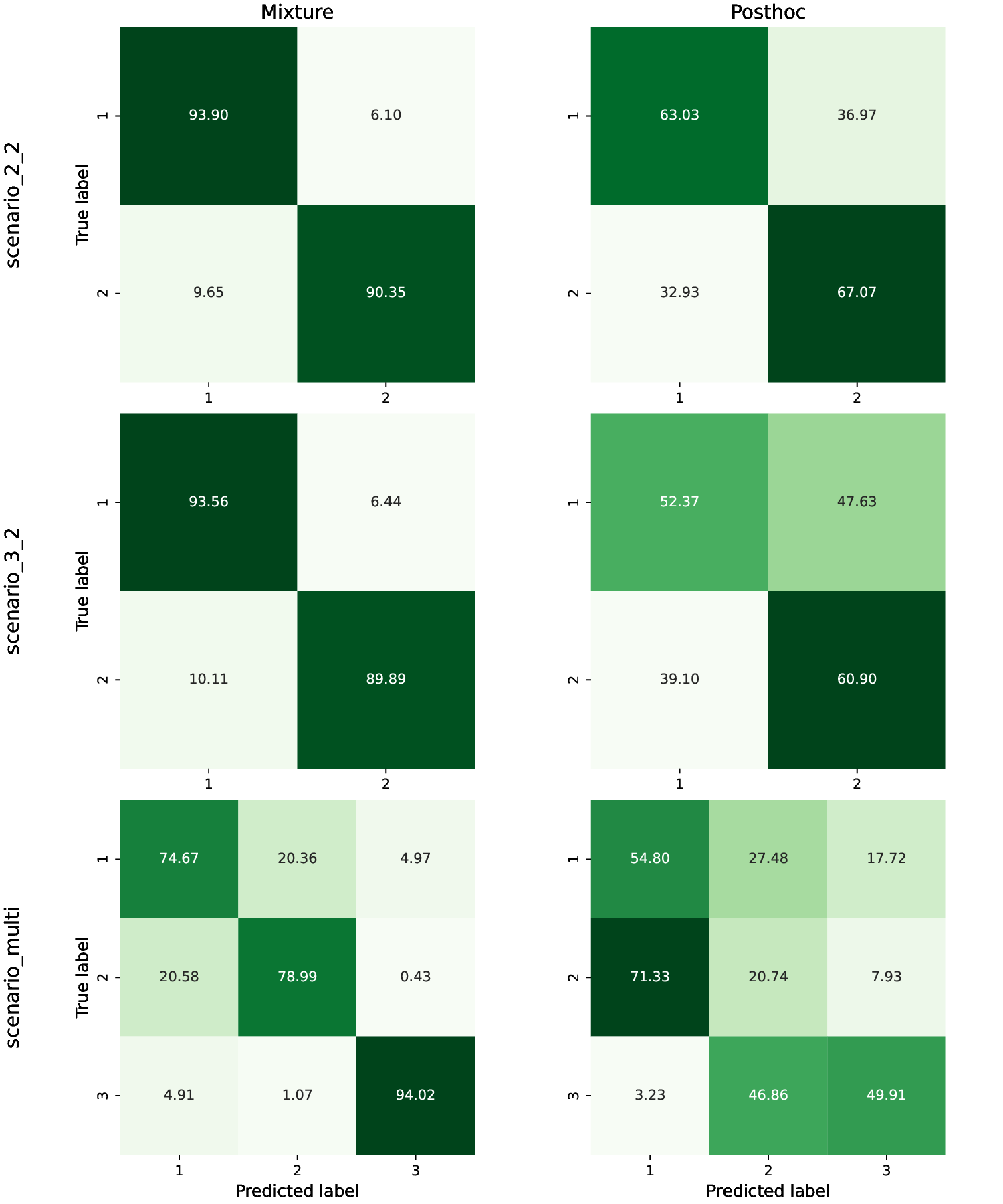}}
\caption{Confusion matrices illustrating the classification performance of the mixture model (left) and the post hoc classification (right) across the three simulated scenarios.
Each cell indicates the proportion of individuals assigned to each predicted cluster relative to their true cluster membership.
Diagonal elements represent the recall for each true cluster. Scenario\_2\_2 refers to the scenario with two scores and two clusters; Scenario\_3\_2 refers to the scenario with three scores and two clusters; Scenario\_multi refers to the scenario with six scores and three clusters.
\label{fig: classification}}
\end{figure}

Figure~\ref{fig: classification} displays the confusion matrices obtained by the mixture model and the post hoc classification, using the correct number of clusters, across the three scenarios.
Overall, the mixture model consistently outperformed the post hoc classification in all cases.
In the first two scenarios, the mixture model successfully recovered the true labels with recall rates exceeding 90\%.
In contrast, the post hoc method achieved only moderate performance, with rates of approximately 63\% and 67\% in \textit{scenario\_2\_2}, and 52\% and 60\% in \textit{scenario\_3\_2}, for the first and second cluster, respectively.
In the third and most complex scenario, the mixture model exhibited robust performance, accurately recovering the fast-progressing cluster with a rate of 94\%, and reliably identifying the first and second clusters with rates of 75\% and 79\%, respectively.
Conversely, the post hoc classification failed to achieve satisfactory separation between clusters. 
It correctly identified the first (average) cluster for only 55\% of individuals, while merely 21\% of individuals were correctly classified to the second (slow-progressing) cluster, with 71\% misassigned to the first cluster.
Similarly, the third (fast-progressing) cluster was recovered with a rate of 50\%, while 47\% of its individuals were incorrectly attributed to the second cluster.

Table~\ref{tab: classification_metrics} summarizes the classification performance of the mixture and post hoc approaches across the three simulation scenarios. 
In the simpler two-cluster scenarios (\textit{scenario\_2\_2 }and \textit{scenario\_3\_2}), the mixture method achieved high overall accuracy (0.91, 95\% CI 0.89–0.93), whereas the post hoc approach showed substantially lower accuracy (0.65 and 0.57, respectively). 
Similar trends were observed for cluster-specific recall and precision, with the mixture method yielding values above 0.85 for most clusters, while post hoc performance was markedly lower, particularly for the first cluster.
In the more complex scenario (\textit{scenario\_multi}), both methods showed reduced performance. 
However, the mixture approach maintained considerably higher accuracy (0.81 vs. 0.42) and higher recall and precision across all three clusters compared to the post hoc approach.
Cluster-specific performance was markedly higher for the mixture model, particularly for the third fast progressing cluster, where recall and precision reached 0.94 and 0.92, respectively. 
Performance for the first and second clusters remained adequate, with both metrics exceeding 0.75. 
In contrast, the post hoc approach performed poorly for the second cluster, with recall and precision of 0.21 and 0.17, respectively, while performance for the remaining clusters remained moderate, ranging from 0.47 to 0.56 for these metrics.
For the third scenario, limiting the analysis to the 932 datasets that favored the three-cluster solution produced almost identical outcomes, with performance metrics varying by no more than 0.01\% (results not shown).

\subsection{Parameter estimation}

For clarity, superscripts denote cluster membership, while subscripts on the $w$ coefficients index the scores.
Table~\ref{tab: estimation_scenario_2_2} reports the parameter estimation results for the \textit{scenario\_2\_2}.
Overall, the mixture model outperformed the post hoc classification across all parameters, yielding estimates that were closer to the true values and characterized by lower RMSEs.
For the cluster probabilities ($\pi^1$, $\pi^2$), the mixture model produced nearly unbiased estimates, whereas the post hoc approach tended to slightly over- or underestimate the true values.
The disease onsets ($\overline{\tau}^1$, $\overline{\tau}^2$), were more accurately recovered by the mixture model, which showed a noticeable reduction in bias and RMSE, relative to the post hoc method.
For the progression rates ($\overline{\xi}^1$, $\overline{\xi}^2$), the mixture model provided nearly unbiased estimates with lower uncertainty, measured by the SE, compared to the post hoc classification.
Finally, the score-specific shifts ($\overline{w_k}^1$, $\overline{w_k}^2$, for $k=1,2$) were recovered more reliably by the mixture model, particularly for the second score $\overline{w_2}$.
More importantly, the post hoc estimates exhibit sign inversions, indicating a reversal in the relative ordering of the scores across clusters. 
Obtaining space-shifts estimates with reversed signs would lead to a misinterpretation, suggesting that the score, which is truly more advanced at disease onset within each cluster appears less advanced. 
This behavior is observed consistently for both scores.

Table~\ref{tab: estimation_scenario_3_2} presents the parameter estimation results for the \textit{scenario\_3\_2} where the clusters are less well separated.
Overall, the mixture model continued to provide more accurate estimates than the post hoc classification, particularly for the cluster proportions and source parameters.
For the cluster proportions ($\pi^1$, $\pi^2$), the mixture model yielded estimates with smaller biases (0.04 and –0.04) compared with the post hoc method (0.12 and –0.12) and substantially lower RMSEs.
For the disease onsets ($\overline{\tau}^1$, $\overline{\tau}^2$), the mixture model achieved slightly smaller biases for the first cluster, while the post hoc classification showed a slightly smaller bias for the second cluster, although the performance remained comparable between methods.
The progression rates ($\overline{\xi}^1$, $\overline{\xi}^2$) were estimated more accurately by the mixture model, with minimal bias and reduced RMSE relative to the post hoc classification.
Finally, the mixture model recovered the score-specific shifts ($\overline{w_k}^1$, $\overline{w_k}^2$, for $k=1,2,3$) better than the post hoc approach, exhibiting null bias in some cases. 
It also performed better than the \textit{scenario\_2\_2}.
Alongside the larger bias, the post hoc classification produced estimates close to zero across all cases.

Table~\ref{tab: estimation_scenario_multi} summarizes parameter recovery for the third more complex scenario (\textit{scenario\_multi}). 
For the cluster proportions ($\pi^c$), the mixture model yielded estimates closer to the true values, with smaller biases and lower RMSEs, particularly for the first and third clusters, whereas the post hoc approach showed substantial estimation error in several cases.
For the onset times ($\overline{\tau}^c$), the mixture model provided more accurate estimates for the first cluster with a bias of -2.89 compared with -8.32 for the post hoc approach.
For the second cluster the post hoc classification overestimated the onset time by 1.44 while the mixture model by 2.02.
The discrepancy was more pronounced for the third cluster, where the post hoc approach showed a minimal bias (–0.16) while the mixture model exhibited a larger bias (–6.17).
The true values indicate identical onset times for the second and third clusters, with a later onset for the first cluster.
The mixture model estimated similar onset times for the first and second clusters, both later than the third cluster, whereas the post hoc classification suggested an ascending sequence, with the first cluster earliest, followed by the third, and then the second cluster.
Additionally, progression rates ($\overline{\xi}^c$) were more accurately estimated by the mixture model, with minimal bias and reduced RMSE relative to the post hoc approach.
From the 18 space-shifts that needed to be estimated, 11 were recovered with smaller bias by the mixture model and 7 by the post hoc approach.
The mixture model seems to have an advantage for scores with larger true shift magnitudes (e.g., $w_4$, $w_5$, and $w_6$), where post hoc estimates tend to be strongly shrunk toward zero and display substantially larger variability.
For scores with near-zero true shifts, both approaches yield small biases, although post hoc classification remains associated with inflated uncertainty.
For RMSE, the mixture approach shows consistently better performance than post hoc classification, which is mainly driven by larger standard errors.
Overall, these results indicate that the mixture model more reliably captures score-specific heterogeneity across clusters, particularly when shifts are moderate to large.
Finally, in this scenario, restricting the evaluation to the 932 datasets for which the three-cluster solution was selected, the results remained virtually identical, with differences in performance metrics of at most 0.01\% (results not shown).

\section{Application to real data}

CADASIL is caused by cysteine missense pathogenic variants in one of the 34 epidermal growth factor-like repeat (EGFr) domains of the NOTCH3 protein.
The clinical spectrum of CADASIL is wide and includes attacks of migraine with aura, stroke, mood disturbances, diverse neuropsychiatric symptoms, cognitive impairment ranging from executive dysfunction to severe dementia, and motor disability~(\cite{Chabriat2009}).
Drawing inspiration from the recent study showing that these manifestations may vary in different groups of patients~(\cite{Kaisaridi2025}) we wanted to analyze the subgroups of similar progression in terms of motor disability and memory performance.
The motor disability was assessed using the modified Rankin Scale and the memory performance using the Total Free Recall score from the Grober and Buschke procedure~(\cite{Epelbaum2011}).

We analyzed a longitudinal dataset of CADASIL patients recruited at the French National Referral Centre CERVCO.
Participants were included if they had at least two visits with available assessments for either the modified Rankin Scale or the Total Free Recall score.
The final analysis comprised 419 patients, corresponding to a total of 2,072 observations.
Across all visits, 12 values were missing for the Rankin score and 507 for the Total Libre score. 
The mean age at baseline was 52 years (standard deviation: 12 years). 
Patients had a median of 4 visits (interquartile range: 3–7), and the average follow-up duration was 7.5 years (standard deviation: 5 years), ranging from 6 months to 20 years.

\begin{figure}[!t]
\centerline{\includegraphics[width=\textwidth]{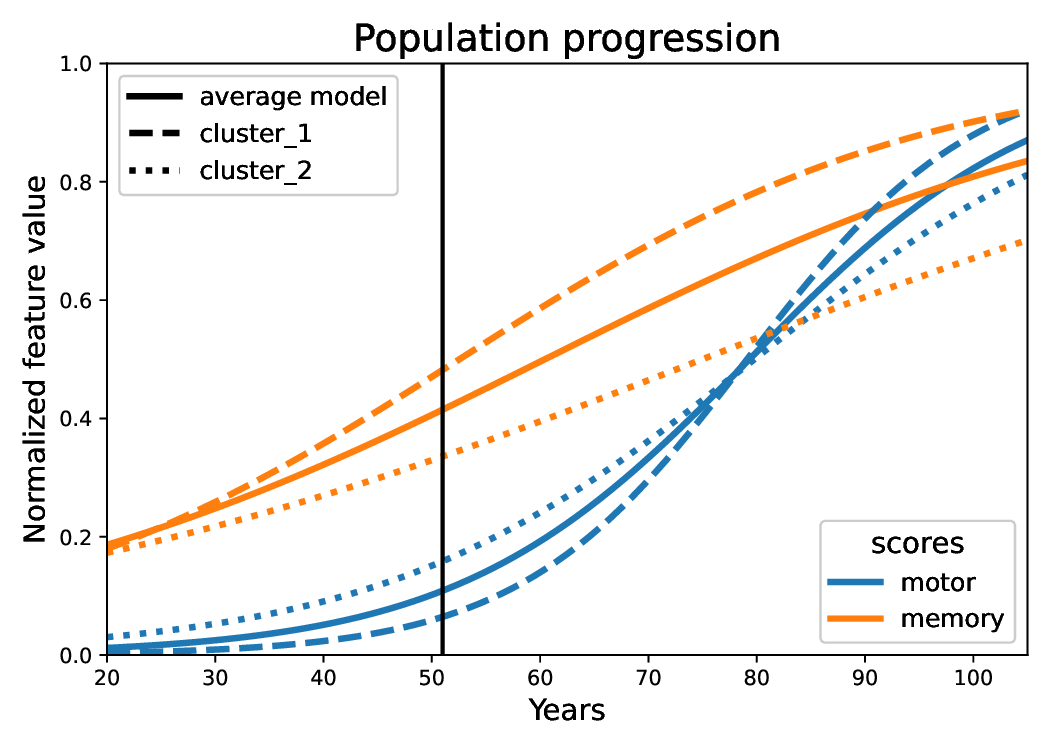}}
\caption{Modeled disease trajectories by cluster and overall average.
Modeled trajectories of disease progression are shown for the two identified clusters and the overall average model. 
The solid lines represent the average model, the dashed lines correspond to Cluster 1, and the dotted lines to Cluster 2. 
The blue curves depict the evolution of the motor score, while the orange curves represent the memory score. 
The vertical solid black line indicates the time of the average disease onset.
\label{fig: mean_traj}}
\end{figure}

We fitted mixture Disease Course Mapping models, implemented in the python library Leaspy, with two, three, and four clusters and computed the corresponding Integrated Completed Likelihood (ICL) values. 
The approximate ICL values were –6970, –6950, and –6925, respectively. 
According to this criterion, the two-cluster model provided the best fit to the data.
The disease trajectory for the two clusters and the average population are depicted in figure \ref{fig: mean_traj}.

The first cluster, comprising 44\% of the patients, was characterized by the parameter set 
$(\tau, \xi, s) = (55.16, 0.28,−2.13)$
The remaining 56\% of individuals formed a second cluster with corresponding parameter values of  $(48.02,−0.22,1.66)$.
To facilitate interpretation, we derived the space-shift parameters $w_k$ from the sources $s$ using leaspy's matrix decomposition. 
This representation allows for a clearer interpretation of the spatial profiles associated with each score and each group.
Accordingly, the parameters $(\tau^c, \xi^c, w_{\text{motor}}^c, w_{\text{memory}}^c)$  were estimated as $(55.16,0.28,−0.03,0.12)$ for the first cluster and $(48.02,−0.22,0.02,−0.09)$ for the second cluster. 
These results suggest that the first cluster corresponds to patients with a later disease onset but a faster rate of progression, characterized by a more advanced memory score and a less advanced motor score at the average disease onset time. 
Conversely, the second cluster represents patients with an earlier disease onset but a slower rate of progression, exhibiting the opposite spatial profile.
Entropy was computed to quantify the uncertainty of cluster membership assignments, reflecting the degree of overlap among clusters. 
The normalized entropy is a standardized measure bounded between 0 and 1, where values approaching 0 denote clear and well-separated clusters, while values near 1 indicate substantial ambiguity in cluster allocation~(\cite{Celeux1996}).
The normalized entropy value is equal to 0.07, which indicates very good cluster separation.

A representation of cluster repartition according to the joint distribution of their individual parameters and the marginal distributions for each individual parameter across clusters can be found in the supplementary material.

\section{Discussion}

We proposed a probabilistic mixture extension of the Disease Course Mapping model capable of handling multivariate scenarios.
The geodesic formulation ensures that smooth, joint evolution of clinical scores, without discontinuities, while parallel transport allows us transition from the population-average trajectory to individual trajectories, through three interpretable subject-specific parameters.
Parameter estimation via the MCMC-SAEM algorithm provides a robust alternative to deterministic optimization by avoiding linearization and efficiently handling high-dimensional latent effects~(\cite{Panhard2008}).
Extending this framework to a mixture setting is a valuable tool for studying disease subtypes.

Simulation studies demonstrated strong robustness.
Model selection using the ICL criterion correctly recovered the true three-cluster structure in over 93\% of cases and estimation and classification performance remained stable even under cluster-number misspecification.
Globally, the mixture model consistently outperformed post hoc classification in recovering both cluster structure and parameters.
Classification accuracy remained high even when clusters overlapped, in contrast to the marked deterioration observed for post-hoc assignment.
This aligns with previous studies showing the superiority of probabilistic mixture models in noisy or overlapping settings~(\cite{McLachlan2000, Fraley2002}).
In a more complex scenario involving three clusters (\textit{scenario\_multi}), the mixture model recovered the fast-progressing cluster with high accuracy (94\%) and identified the remaining two with performance exceeding 75\%. 
This discrepancy aligns with previous findings that leaspy effectively detects fast progressors~(\cite{Maheux2023}), and other studies showing higher accuracy for clusters that are more distinct in feature space~(\cite{Young2021}). 
The first cluster reflects an average profile in terms of progression rate with intermediate space-shift values, which may partially account for the model’s limited ability to distinguish patients in the first two clusters.
Similar patterns have been reported in disease subtyping, where clusters with intermediate profiles are more likely to be misclassified~(\cite{Vidal2014, Marquand2016}).
In contrast, post hoc classification failed to recover the correct clustering structure, highlighting the importance of explicitly modeling cluster-specific distributions and progression trajectories~(\cite{Bzdok2017}).

Furthermore, the mixture model globally outperforms the post hoc classification in recovering the parameters that define the clusters. 
In the simplest scenario (\textit{scenario\_2\_2}) the mixture model yielded smaller bias, lower RMSE, and reduced uncertainty. 
It preserved relative differences in onset times and more reliably estimated score-specific shifts, underscoring its ability to capture heterogeneity in spatial progression.
In a more challenging configuration (\textit{scenario\_3\_2}), with more intertwined clusters, the mixture model provided substantially less biased estimates for all parameters except for the disease onset parameter ($\tau$) in the second cluster, although it preserves the relative difference between clusters, whereas the post hoc approach produces nearly identical estimates.
These findings are consistent with prior work showing improved parameter recovery under latent‑class compared to hard‑assignment approaches~(\cite{Vandernest2020}).
With respect to spatial progression post hoc classification produces near-zero estimates, suggesting a failure to detect meaningful spatial variation.
Interestingly, this added complexity appears to have a positive effect on the mixture model, which shows an improvement in bias compared to \textit{scenario\_2\_2}. 
The inclusion of an additional, similarly progressing score provides an additional reference point, thereby enhancing the calibration of these spatial parameters.
In the most complex scenario, involving six scores and three clusters, the mixture model continued to estimate probability parameters, progression rates, and score-specific shifts with lower bias and standard errors.
While it was not uniformly superior for every parameter, it more faithfully captured cluster ordering and the direction of progression. 
post hoc classification frequently distorted temporal ordering and underestimated pronounced spatial shifts, particularly for larger true $w$ values, underestimating meaningful heterogeneity.

A key advantage of this method is its strong performance in scenarios with overlapping clusters.
By jointly estimating cluster memberships and cluster-specific parameters, it accounts for uncertainty in class assignment and avoids the biases introduced by post hoc classification.
Unlike cross-sectional clustering approaches~(\cite{Young2018}), our longitudinal formulation clusters patients based on disease dynamics, capturing the spatial patterns that govern the order of symptom appearance and the temporal variations in disease onset and progression rate.
The time reparametrization scheme, which is absent in other latent class models, enables interpretation in terms of disease age rather than chronological age, which is particularly relevant for slowly progressive disorders.
Importantly, relative ordering of temporal parameters remains meaningful at the mixture extension, supporting clinical interpretability: a cluster with an onset value of 50 can be confidently interpreted as having an earlier disease onset than a cluster with an onset value of 60.
Therefore, the mixture model’s consistent recovery of the relative differences between cluster parameters further underscores the robustness of our method.
We did not compare our method with alternatives such as \texttt{lcmm}~(\cite{Proust2017}), as it is formulated on chronological time, whereas our approach explicitly models a latent disease timeline, separating disease stage variability from variability due to differences in observation timing. 
Consequently, the two frameworks address different inferential targets, and a direct comparison would not isolate methodological performance but rather reflect the differences in the underlying definition of time and the progression process being modeled.
While alternative hierarchical mixture approaches model both individual- and population-level mixtures~(\cite{Viani2025}), our formulation preserves the global trajectory, facilitating interpretation of population trends.
These assets ensure clear identification of cluster centers and allow direct interpretation as distinct disease phenotypes as shown in the real data application.

Nonetheless, limitations should be acknowledged. 
The number of clusters is determined a posteriori, although this remains standard practice in finite mixture modeling~(\cite{Proust2017}). 
Bayesian nonparametric alternatives, such as Dirichlet-based approaches~(\cite{Rouanet2024}), could allow simultaneous estimation of cluster number. 
Additionally, inferred trajectories depend on the selected clinical measures, whose scaling properties may bias estimated progression patterns. 
Careful score selection and potential integration of formal variable selection procedures could further enhance robustness and interpretability.

\section{Conclusion}
Overall, our results demonstrate that the proposed mixture model provides a robust and interpretable framework for capturing spatiotemporal heterogeneity in disease progression, offering a valuable tool for patient stratification and the study of complex, multivariate clinical trajectories.
The approach proves particularly useful in scenarios where clusters are overlapping, yielding more reliable and stable estimations than existing post hoc classification solutions.




%
\bibliographystyle{biom} 
\bibliography{biomsample_bib}






\section*{Supporting Information}

Supplementary material is available with this paper at the Biometrics website on Wiley Online
Library.\vspace*{-8pt}





\label{lastpage}

\clearpage

\begin{sidewaystable}[!t]%
\centering %
\caption{Mean values of the distributions defining the simulation parameters for each cluster across scenarios. For each scenario, values in parentheses indicate the parameters associated with the respective cluster. Scenario\_2\_2 refers to the scenario with two scores and two clusters; Scenario\_3\_2 refers to the scenario with three scores and two clusters; Scenario\_multi refers to the scenario with six scores and three clusters.\label{tab: simulation_values}}%
\begin{tabular*}{\textwidth}{@{\extracolsep\fill}lllll@{\extracolsep\fill}}
\toprule
\textbf{Parameter} & \textbf{Interpretation} & \textbf{scenario\_2\_2} & \textbf{scenario\_3\_2} & \textbf{scenario\_multi} \\
\midrule
$\pi$ & probability & (0.40, 0.60) & (0.40, 0.60) & (0.40, 0.35, 0.25) \\
$\overline{\tau}$ & disease onset &(50.00, 40.00) & (56.00, 53.00) & (70.00, 65.00, 65.00) \\
$\overline{\xi}$ & progression rate & (-0.30, 0.20) & (0.30, -0.20) & (0, -0.40, 0.5) \\
$w_1$ & score-specific shift & (-0.02, 0.02) & (-0.06, 0.05) & (0.00, 0.00, - 0.01) \\
$w_2$ & score-specific shift & (0.11, -0.11) & (0.07, -0.06) & (-0.01, 0.00, -0.01) \\
$w_3$ & score-specific shift & - & (0.01, -0.01) & (-0.01, -0.01, -0.02) \\
$w_4$ & score-specific shift & - & - & (0.05, 0.00, 0.10) \\
$w_5$ & score-specific shift & - & - & (0.11, -0.03, 0.27) \\
$w_6$ & score-specific shift & - & - & (0.15, 0.03, 0.24)\\
\bottomrule
\end{tabular*}
\end{sidewaystable}

\begin{sidewaystable}
\centering
\caption{Classification performance metrics. For each scenario the achieved overall accuracy and the per-class recall and precision are reported for the mixture model and the post hoc classification. Scenario\_2\_2 refers to the scenario with two scores and two clusters; Scenario\_3\_2 refers to the scenario with three scores and two clusters; Scenario\_multi refers to the scenario with six scores and three clusters.\label{tab: classification_metrics}}
\begin{tabular*}{\textwidth}{@{\extracolsep\fill}lrllllll@{}}
\toprule
&&\multicolumn{2}{@{}l}{\textbf{scenario\_2\_2}} & \multicolumn{2}{@{}l}{\textbf{scenario\_3\_2}} & \multicolumn{2}{@{}l}{\textbf{scenario\_multi}} \\
\textbf{metric} & & \textbf{mixture}  & {\textbf{post hoc}}  & \textbf{mixture} & {\textbf{post hoc}}  & \textbf{mixture} & {\textbf{post hoc}}     \\
\midrule
Accuracy && 0.91 (0.89-0.93) & 0.65 (0.62-0.68) & 0.91 (0.89-0.93) & 0.57 (0.54-0.60) & 0.81 (0.78-0.83) & 0.42 (0.39-0.45) \\
\midrule
Recall & cluster 1 & 0.94 (0.92-0.95) & 0.63 (0.60-0.66) & 0.94 (0.92-0.95) & 0.52 (0.50-0.55) & 0.75 (0.72-0.77) & 0.55 (0.52-0.58)\\
    & 2 & 0.90 (0.88-0.92) & 0.67 (0.64-0.70) & 0.90 (0.88-0.92) & 0.61 (0.58-0.64) & 0.79 (0.76-0.81) & 0.21 (0.18-0.23)\\
    & 3 & - & - & - & - & 0.94 (0.92-0.95) & 0.50 (0.47-0.53)\\
\midrule
Precision & cluster 1 & 0.87 (0.84-0.89) & 0.54 (0.51-0.58) & 0.86 (0.84-0.88) & 0.47 (0.44-0.50) & 0.78 (0.75-0.80) & 0.47 (0.44-0.50)\\
    & 2 & 0.96 (0.94-0.97) & 0.75 (0.72-0.77) & 0.96 (0.94-0.97) & 0.66 (0.63-0.69) & 0.76 (0.74-0.79) & 0.17 (0.15-0.20)\\
     & 3 & - & - & - & - & 0.92 (0.90-0.93) & 0.56 (0.53-0.59)\\
\bottomrule
\end{tabular*}
\end{sidewaystable}

\begin{center}
\begin{sidewaystable}[!t]
\caption{Summary of parameter recovery in scenario\_2\_2. For each parameter, the estimates by the mixture model and post hoc classification are compared in terms of bias, standard error (SE), and root-mean-square error (RMSE). The true values are also reported. For clarity, superscripts denote cluster membership, while subscripts on the $w$ coefficients index the scores. Scenario\_2\_2 refers to the scenario with two scores and two clusters. The values with the smallest biases are marked with * for each parameter. $\pi^c$ denote the cluster probabilities, $\tau^c$ the disease onsets, $\xi^c$ the progression rates and $w_k^c$ the score specific shifts.
\label{tab: estimation_scenario_2_2}}
\begin{tabular*}{\textwidth}{@{\extracolsep\fill}cccccccccc@{}}
\toprule
&&\multicolumn{2}{@{}c}{\textbf{Estimate}} & \multicolumn{2}{@{}c}{\textbf{Bias}} & \multicolumn{2}{@{}c}{\textbf{SE}} & \multicolumn{2}{@{}c}{\textbf{RMSE}} \\
\textbf{parameter} & \textbf{true value} & \textbf{mixture}  & {\textbf{post hoc}}  & \textbf{mixture} & {\textbf{post hoc}}  & \textbf{mixture} & {\textbf{post hoc}} & \textbf{mixture} & {\textbf{post hoc}} \\
\midrule
$\pi^1$ & 0.40 & 0.43 & 0.45 & 0.03* & 0.05 & 0.01 & 0.05 & 0.04 & 0.07 \\
$\pi^2$ & 0.60 & 0.57 & 0.55 & -0.03* & -0.05 & 0.01 & 0.05 & 0.04 & 0.07 \\
\midrule
$\overline{\tau}^1$ & 50.00 & 49.87 & 48.37 & -0.13* & -1.63 & 1.64 & 3.97 & 1.64 & 4.29 \\
$\overline{\tau}^2$ & 40.00 & 39.76 & 41.25 & -0.24* & 1.25 & 1.38 & 3.81 & 1.40 & 4.01 \\
\midrule
$\overline{\xi}^1$ & -0.30 & -0.20 & -0.15 & 0.10* & 0.15 & 0.04 & 0.13 & 0.11 & 0.20 \\
$\overline{\xi}^2$ & 0.20 & 0.15 & 0.10 & -0.05* & -0.10 & 0.03 & 0.13 & 0.06 & 0.16 \\
\midrule
$\overline{w_1}^1$ & -0.02 & -0.05 & 0.02 & -0.03* & 0.04 & 0.01 & 0.03 & 0.03 & 0.05 \\
$\overline{w_1}^2$ & 0.02 & 0.04 & -0.02 & 0.02* & -0.04 & 0.01 & 0.02 & 0.02 & 0.04 \\
$\overline{w_2}^1$ & 0.11 & 0.06 & -0.02 & -0.06* & -0.13 & 0.01 & 0.02 & 0.06 & 0.13 \\
$\overline{w_2}^2$ & -0.11 & -0.04 & 0.02 & 0.06* & 0.13 & 0.01 & 0.02 & 0.06 & 0.13 \\
\bottomrule
\end{tabular*}
\end{sidewaystable}
\end{center}

\begin{center}
\begin{sidewaystable}[!t]
\caption{Summary of parameter recovery in \textit{scenario\_3\_2}. For each parameter, the estimates by the mixture model and post hoc classification are compared in terms of bias, standard error (SE), and root-mean-square error (RMSE). The true values are also reported. For clarity, superscripts denote cluster membership, while subscripts on the $w$ coefficients index the scores. Scenario\_3\_2 refers to the scenario with three scores and two clusters. The values with the smallest biases are marked with * for each parameter. $\pi^c$ denote the cluster probabilities, $\tau^c$ the disease onsets, $\xi^c$ the progression rates and $w_k^c$ the score specific shifts.
\label{tab: estimation_scenario_3_2}}
\begin{tabular*}{\textwidth}{@{\extracolsep\fill}cccccccccc@{}}
\toprule
&&\multicolumn{2}{@{}c}{\textbf{Estimate}} & \multicolumn{2}{@{}c}{\textbf{Bias}} & \multicolumn{2}{@{}c}{\textbf{SE}} & \multicolumn{2}{@{}c}{\textbf{RMSE}} \\
\textbf{parameter} & \textbf{true value} & \textbf{mixture}  & {\textbf{post hoc}}  & \textbf{mixture} & {\textbf{post hoc}}  & \textbf{mixture} & {\textbf{post hoc}} & \textbf{mixture} & {\textbf{post hoc}} \\
\midrule
$\pi^1$ & 0.40 & 0.44 & 0.52 & 0.04* & 0.12 & 0.01 & 0.06 & 0.04 & 0.14 \\
$\pi^2$ & 0.60 & 0.57 & 0.48 & -0.04* & -0.12 & 0.01 & 0.06 & 0.04 & 0.14 \\
\midrule
$\overline{\tau}^1$ & 56.00 & 53.31 & 52.34 & -2.68* & -3.66 & 0.57 & 0.83 & 2.76 & 3.75 \\
$\overline{\tau}^2$ & 53.00 & 51.94 & 52.91 & -1.06 & -0.09* & 0.59 & 0.87 & 1.21 & 0.88 \\
\midrule
$\overline{\xi}^1$ & 0.30 & 0.24 & -0.05 & -0.05* & -0.35 & 0.03 & 0.20 & 0.06 & 0.41 \\
$\overline{\xi}^2$ & -0.20 & -0.19 & 0.11 & 0.01* & 0.31 & 0.02 & 0.20 & 0.03 & 0.37 \\
\midrule
$\overline{w_1}^1$ & -0.06 & -0.06 & -0.01 & 0.00* & 0.05 & 0.00 & 0.01 & 0.00 & 0.05 \\
$\overline{w_1}^2$ & 0.05 & 0.04 & 0.00 & -0.01* & -0.05 & 0.00 & 0.01 & 0.01 & 0.05 \\
$\overline{w_2}^1$ & 0.07 & 0.05 & 0.01 & -0.02* & -0.06 & 0.00 & 0.01 & 0.02 & 0.07 \\
$\overline{w_2}^2$ & -0.06 & -0.03 & 0.00 & 0.03* & 0.06 & 0.00 & 0.01 & 0.02 & 0.06 \\
$\overline{w_3}^1$ & 0.01 & 0.01 & 0.00 & 0.00* & -0.01 & 0.00 & 0.00 & 0.00 & 0.01 \\
$\overline{w_3}^2$ & -0.01 & -0.01 & 0.00 & 0.00* & 0.01 & 0.00 & 0.00 & 0.00 & 0.01 \\
\bottomrule
\end{tabular*}
\end{sidewaystable}
\end{center}

\begin{center}
\begin{sidewaystable}[!t]
\caption{Summary of parameter recovery in \textit{scenario\_multi}. For each parameter, the estimates by the mixture model and post hoc classification are compared in terms of bias, standard error (SE), and root-mean-square error (RMSE). The true values are also reported. The values with the smallest biases are marked with *. When the bias is equal for the two approaches (marked with **) the comparison in done also for the SE. For clarity, superscripts denote cluster membership, while subscripts on the $w$ coefficients index the scores. Scenario\_multi refers to the scenario with six scores and three clusters. $\pi^c$ denote the cluster probabilities, $\tau^c$ the disease onsets, $\xi^c$ the progression rates and $w_k^c$ the score specific shifts.
\label{tab: estimation_scenario_multi}}
\begin{tabular*}{\textwidth}{@{\extracolsep\fill}cccccccccc@{}}
\toprule
&&\multicolumn{2}{@{}c}{\textbf{Estimate}} & \multicolumn{2}{@{}c}{\textbf{Bias}} & \multicolumn{2}{@{}c}{\textbf{SE}} & \multicolumn{2}{@{}c}{\textbf{RMSE}} \\
\textbf{parameter} & \textbf{true value} & \textbf{mixture}  & {\textbf{post hoc}}  & \textbf{mixture} & {\textbf{post hoc}}  & \textbf{mixture} & {\textbf{post hoc}} & \textbf{mixture} & {\textbf{post hoc}} \\
\midrule
$\pi^1$ & 0.40 & 0.38 & 0.30 & -0.02* & -0.10 & 0.02 & 0.06 & 0.03 & 0.12 \\
$\pi^2$ & 0.35 & 0.36 & 0.36 & 0.01** & 0.01** & 0.02* & 0.03 & 0.002 & 0.04  \\
$\pi^3$ & 0.25 & 0.26 & 0.34 & 0.01* & 0.09 & 0.02 & 0.05 & 0.02 & 0.11 \\
\midrule
$\overline{\tau}^1$ & 70.00 & 67.11 & 61.68 & -2.89* & -8.32 & 0.92 & 4.03 & 3.04 & 9.24 \\
$\overline{\tau}^2$ & 65.00 & 67.02 & 66.44 & 2.02 & 1.44* & 1.19 & 2.40 & 2.34 & 2.80 \\
$\overline{\tau}^3$ & 65.00 & 58.83 & 64.84 & -6.17 & -0.16* & 1.00 & 3.84 & 6.25 & 3.84 \\
\midrule
$\overline{\xi}^1$ & 0.00 & -0.02 & 0.34 & -0.02* & 0.34 & 0.15 & 0.44 & 0.15 & 0.56 \\
$\overline{\xi}^2$ & -0.40 & -0.43 & -0.18 & -0.03* & 0.23 & 0.17 & 0.34 & 0.17 & 0.41 \\
$\overline{\xi}^3$ & 0.50 & 0.64 & 0.02 & 0.14* & -0.48 & 0.09 & 0.44 & 0.17 & 0.65 \\
\midrule
$\overline{w_1}^1$ & 0.00 & 0.01 & 0.02 & 0.01* & 0.02 & 0.01 & 0.09 & 0.01 & 0.09 \\
$\overline{w_1}^2$ & 0.00 & 0.03 & -0.01 & 0.03 & -0.01* & 0.01 & 0.07 & 0.03 & 0.07 \\
$\overline{w_1}^3$ & -0.01 & -0.04 & 0.00 & -0.03 & 0.01* & 0.01 & 0.07 & 0.03 & 0.08 \\

$\overline{w_2}^1$ & -0.01 & 0.01 & 0.02 & 0.02* & 0.03 & 0.01 & 0.10 & 0.02 & 0.11 \\
$\overline{w_2}^2$ & 0.00 & 0.03 & -0.01 & 0.03 & -0.01* & 0.01 & 0.08 & 0.03 & 0.08 \\
$\overline{w_2}^3$ & -0.01 & -0.05 & 0.00 & -0.04 & 0.01* & 0.01 & 0.09 & 0.04 & 0.09 \\

$\overline{w_3}^1$ & -0.01 & 0.01 & 0.02 & 0.02* & 0.03 & 0.01 & 0.08 & 0.02 & 0.09 \\
$\overline{w_3}^2$ & -0.01 & 0.01 & -0.01 & 0.02 & 0.00* & 0.01 & 0.06 & 0.02 & 0.06 \\
$\overline{w_3}^3$ & -0.02 & -0.03 & 0.00 & -0.01* & 0.02 & 0.01 & 0.07 & 0.02 & 0.07 \\

$\overline{w_4}^1$ & 0.05 & -0.03 & -0.06 & -0.08* & -0.11 & 0.02 & 0.25 & 0.08 & 0.27\\
$\overline{w_4}^2$ & 0.00 & -0.06 & 0.04 & -0.06 & 0.03* & 0.02 & 0.18 & 0.07 & 0.19 \\
$\overline{w_4}^3$ & 0.10 & 0.12 & 0.00 & 0.02* & -0.10 & 0.02 & 0.21 & 0.02 & 0.23 \\

$\overline{w_5}^1$ & 0.11 & -0.01 & -0.05 & -0.12* & -0.16 & 0.03 & 0.26 & 0.13 & 0.31 \\
$\overline{w_5}^2$ & -0.03 & -0.09 & 0.03 & -0.06** & 0.06** & 0.03* & 0.20 & 0.07 & 0.20 \\
$\overline{w_5}^3$ & 0.27 & 0.13 & 0.01 & -0.14* & -0.27 & 0.02 & 0.22 & 0.15 & 0.35 \\

$\overline{w_6}^1$ & 0.15 & -0.05 & -0.09 & -0.20* & -0.24 & 0.03 & 0.38 & 0.20 & 0.45 \\
$\overline{w_6}^2$ & 0.03 & -0.07 & 0.06 & -0.10 & 0.03* & 0.03 & 0.28 & 0.10 & 0.28 \\
$\overline{w_6}^3$ & 0.24 & 0.15 & -0.01 & -0.09* & -0.25 & 0.02 & 0.31 & 0.09 & 0.40 \\
\bottomrule
\end{tabular*}
\end{sidewaystable}
\end{center}

\end{document}